\documentclass[prl,amsmath,amssymb,twocolumn,superscriptaddress]{revtex4-1}
\usepackage{graphicx,bm,color,appendix}
\usepackage[colorlinks,bookmarks=true,citecolor=blue,linkcolor=blue,urlcolor=blue]{hyperref}

\newcommand{\mG}{\mathcal{G}}
\newcommand{\kl}{K\"ahler~}
\newcommand{\mr}{moir\'e~}

\newcommand{\mT}{\mathcal{T}~}
\newcommand{\mP}{\mathcal{P}}
\newcommand{\mC}{\mathcal{C}}
\newcommand{\mD}{\mathcal{D}}
\newcommand{\mK}{\mathcal{K}}
\newcommand{\mI}{\mathcal{I}}

\begin{document}
\title{Hierarchy of Ideal Flatbands in Chiral Twisted Multilayer Graphene Models}

\author{Jie Wang}
\email{jiewang@flatironinstitute.org}
\affiliation{Center for Computational Quantum Physics, Flatiron Institute, 162 5th Avenue, New York, New York 10010, USA}
\author{Zhao Liu}
\email{zhaol@zju.edu.cn}
\affiliation{Zhejiang Institute of Modern Physics, Zhejiang University, Hangzhou 310027, China}

\begin{abstract}
We propose models of twisted multilayer graphene that exhibit exactly flat Bloch bands with arbitrary Chern numbers and ideal band geometries. The models are constructed by twisting two sheets of Bernal-stacked multiple graphene layers with only inter-sublattice couplings. Analytically we show that flatband wavefunctions in these models exhibit a momentum space holomorphic character, leading to ideal band geometries. We also explicitly demonstrate a generic ``wavefunction exchange'' mechanism that generates the high Chern numbers of these ideal flatbands. The ideal band geometries and high Chern numbers of the flatbands imply the possibility of hosting exotic fractional Chern insulators which do not have analogues in continuum Landau levels. We numerically verify that these exotic fractional Chern insulators are model states for short-range interactions, characterized by exact ground-state degeneracies at zero energy and infinite particle-cut entanglement gaps.
\end{abstract}

\maketitle
\emph{Introduction.---}The intrinsic topological and geometric properties of Bloch wavefunctions are crucial to the interacting phenomena in narrow-band systems such as \mr materials~\cite{Andrei:2020aa,Balents:2020aa,Kennes:2021aa} where the electrons' kinetic energies are quenched. The band topology enriches the possible many-body phase diagram~\cite{PhysRevX.1.021014,PARAMESWARAN2013816,zhao_review}. On the other hand, the band geometry determines the actual stabilities of various many-body states~\cite{Jackson:2015aa}.

As a representative example, twisted bilayer graphene (TBG) has two nearly flatbands of Chern number $\mC=\pm1$ at charge neutrality. Recently, fractional Chern insulators (FCIs)~\cite{PhysRevX.1.021014,PARAMESWARAN2013816,zhao_review} were theoretically predicted and experimentally observed in TBG flatbands~\cite{ZhaoLiu_TBG,Cecile_TBG_Flatband,Cecile_PRL20,FCI_TBG_exp}. One important factor to the stability of FCIs in this system is due to the ideal geometry of the flatbands in the fixed point chiral limit~\cite{Grisha_TBG,Oscar_hiddensym}, where each flatband's Berry curvature $\Omega_{\bm k}$ is non-vanishing and strictly proportional to its Fubini-Study metric $g^{ab}_{\bm k}$ by a constant determinant-one matrix $\omega^{ab}$~\cite{Grisha_TBG2,JieWang_NodalStructure,JieWang_exactlldescription}:
\begin{equation}
g^{ab}_{\bm k} = \frac{1}{2}\omega^{ab}\Omega_{\bm k},\quad \Omega_{\bm k}\neq0 \quad \text{for}\quad\forall\bm k,
\label{idealcond}
\end{equation}
where $a,b=x,y$ labels spatial coordinates. The ideal band geometry Eq.~(\ref{idealcond}) implies the Bloch wavefunctions of the chiral TBG (cTBG) flatbands exhibit a momentum space holomorphic character~\cite{Martin_PositionMomentumDuality,Lee_engineering}, in analogy to the real space holomorphic wavefunction in the conventional lowest Landau level (LLL). Such exact position-momentum duality leads to the existence of model FCIs in the cTBG flatbands as the exact zero-energy ground states of short-range interactions which are stable against the spatial fluctuation of band geometries~\cite{Grisha_TBG2,JieWang_exactlldescription}.

\begin{figure}
\centering
\includegraphics[width=\linewidth]{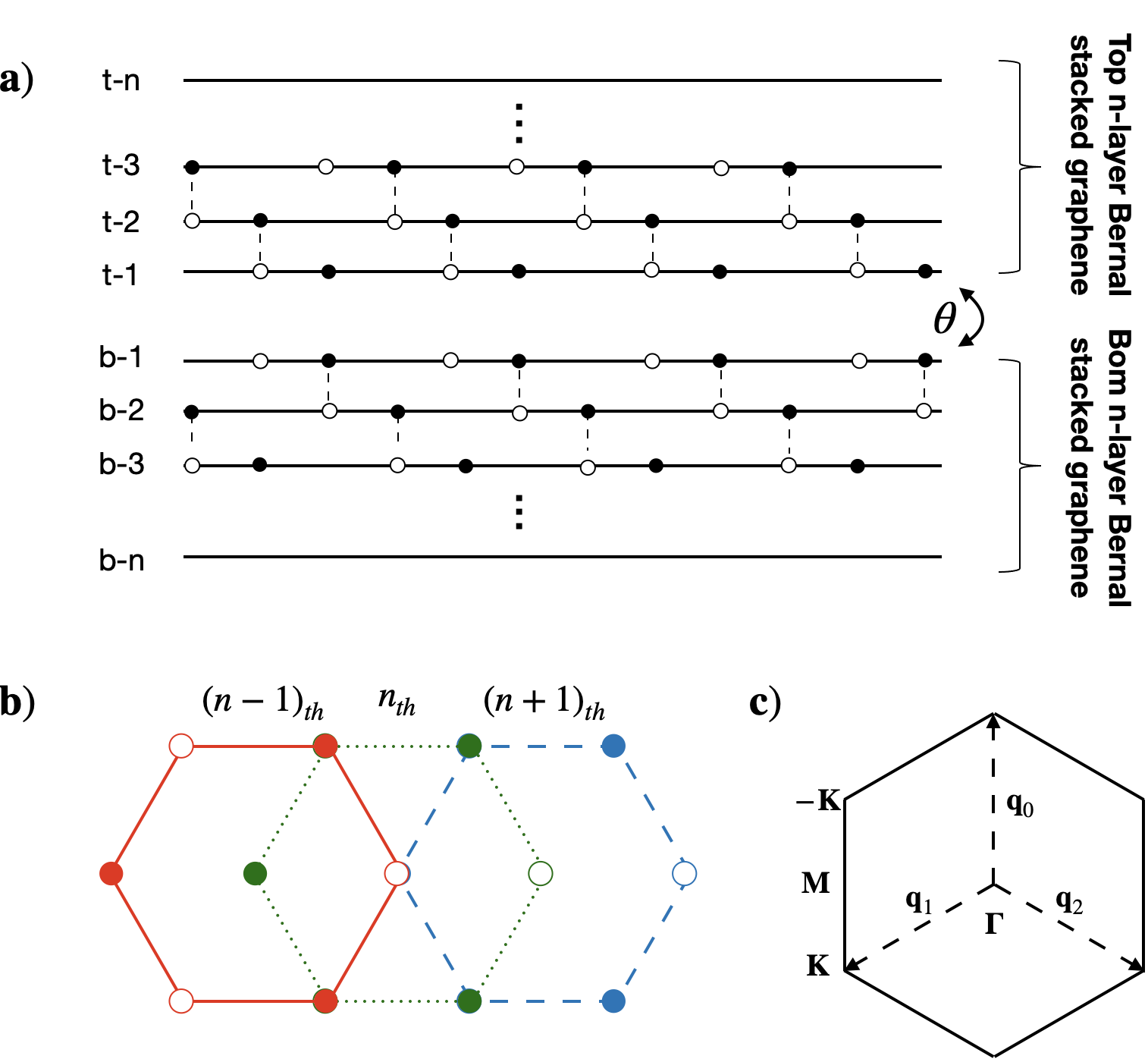}
\caption{(a) Geometry of our $AB-AB$ Bernal stacking multi-layered model, which consists of $n$ layers of Bernal stacked graphene on top and bottom sheets, respectively, with a relative small twisted angle $\theta$ in the middle. Here the solid and empty dots represent the $A$ and $B$ sublattice, respectively. (b) Illustration of the Bernal stacking structure in three consecutive layers. (c) The \mr Brillouin zone and some important momentum points.}
\label{fig:model}
\end{figure}

The ideal flatbands are special cases of the \kl band~\cite{kahlerband1,kahlerband2,kahlerband3} when the \kl structure~\cite{Douglas:2009aa} is spatially constant. There is so far a glaring lack of microscopic models realizing ideal flatbands of high Chern numbers (high-$\mC$). Compared with $|\mC|=1$ bands, high-$\mC$ bands are topologically different~\cite{Trescher2012,Yang2012,2021arXiv210514672K} and may support many-body phases without LL analogues~\cite{Zhao_FCI_highC,Sterdyniak13,Yangle_Modelwf,Moller_FCI_highC,YinghaiWu_HighC,ModelFCI_Zhao,Bart_HighC,Bart_stability,Bart_FQHtransition}. In this work, we fill this void and propose a systematic construction of microscopic models with relevance to \mr materials. Our models are based on two sheets of $n$-layer Bernal stacked graphene which are twisted by a small angle and put in the chiral limit. Our hierarchy scheme starts with cTBG as the parent, and includes the chiral twisted double bilayer graphene (cTDBG) as the next descendant~\cite{Koshino_19PRB,Koshino_20PRB,TDBG_nanoletter,Lee:2019aa,ZhaoLiu_TDBG,Shi:2020aa,Geisenhof:2021aa,PhysRevLett.106.156801,PhysRevB.89.035405}. We show exactly flat bands existing at charge neutrality of our models, and we analytically and mathematically prove their ideal band geometry and exotic band topology. We also numerically show that lattice-specific FCIs without LLL analogues are stable in these ideal flatbands as they appear as the exact zero-energy ground states of short-range interactions, paving the way towards understanding their stability against inhomogeneous band geometries.

\emph{Multilayer Chiral Model.---}We consider two sheets of $n$-layer Bernal stacked graphene twisted by a small angle $\theta$, as illustrated in Fig.~\ref{fig:model}. We focus on a single valley of the system~\cite{Bistritzer12233,Santos,Santos2} and take the chiral limit~\cite{Grisha_TBG} by keeping only the inter-sublattice hopping between adjacent layers, such that the Hamiltonian of our model takes an off-diagonal form in the sublattice basis
\begin{equation}
H_n = \left(\begin{matrix}\Phi^{\dag}&\Xi^{\dag}\end{matrix}\right)\left(\begin{matrix} & \mD_n \\ \mD_n^{\dag} &\end{matrix}\right)\left(\begin{matrix}\Phi\\\Xi\end{matrix}\right),\label{model}
\end{equation}
where the basis $\Phi$ and $\Xi$ are fully sublattice-$A$ and $B$ polarized. We organize $\Phi$ (and $\Xi$) by layers, such that $\Phi=\left(\phi_1,\phi_2,...,\phi_n\right)^T$ where $\phi_i=(\phi_i^b,\phi_i^t)^T$ contains the sublattice-$A$ components of the $i$th layer in the bottom sheet and the $i$th layer in the top sheet (Fig.~\ref{fig:model}). In this basis, we have
\begin{equation}
\mD_{n} = \left(\begin{matrix}\mD_1 & t_1T_+ & \\ t_1T_- & h_D & t_2T_+ & \\ & t_2T_- & h_D & \ddots \\ && \ddots && \\ &&&& t_{n-1}T_+ \\ &&&t_{n-1}T_- & h_D\end{matrix}\right),\nonumber
\end{equation}
where $h_D$ and $\mD_1$ are, respectively, the Dirac Hamiltonian of a freestanding monolayer graphene and the Hamiltonian of cTBG~\cite{Grisha_TBG}, given by
\begin{equation}
h_D = \left(\begin{matrix}-i\partial & \\&-i\partial\end{matrix}\right),\quad\mD_1 = \left(\begin{matrix}-i\partial & U_{-\phi} \\ U^*_{\phi} & -i\partial\end{matrix}\right).\label{h_D}
\end{equation}

Here $U_{\phi} = \alpha\left(e^{-i\bm q_0\cdot\bm r} + e^{i\phi}e^{-i\bm q_1\cdot\bm r} + e^{-i\phi}e^{-i\bm q_2\cdot\bm r}\right)$ with $\phi=2\pi/3$ and $\alpha\propto\sin^{-1}(\theta/2)$~\cite{Grisha_TBG}, $\partial=(\partial_x-i\partial_y)/\sqrt2$, and the momenta $\bm q_{0,1,2}$ are illustrated in Fig.~\ref{fig:model}(c). The twist angle $\theta$ is set as the magic angle of cTBG~\cite{Grisha_TBG}. We only retain the strongest tunneling in the Bernal stacking structure [Fig.~\ref{fig:model}(b)], which couples electrons in the $n$th layer sublattice-$A$ to those in the $(n+1)$th layer sublattice-$B$ by the real tunneling strength $t_n$ [Fig.~\ref{fig:model}(a)]. Under this assumption, the interlayer coupling matrices $T_{\pm}$ are
\begin{equation}
T_+ = \left(\begin{matrix}1&0\\0&0\end{matrix}\right),\quad T_- = \left(\begin{matrix}0&0\\0&1\end{matrix}\right).
\end{equation}

Details of the model Hamiltonian are left to the Supplementary Material (SM).

The model Eq.~(\ref{model}) preserves the translation $\mathcal{V}_{1,2}$ and the threefold rotation $\mC_3$ symmetries, but it breaks the time-reversal $\mT$ and the twofold rotation $\mC_2$ as they interchange valleys~\cite{Mele_TBG_Sym,Po_Symmetry_PRX,Zou_Symmetry_PRB,Zhida_PRL19,Zaletel_PRX20}. The combination $\mC_2\mT$ is also broken by the Bernal-stacking unless $n=1$. Besides the lattice symmetries, the model has two exact emergent symmetries: the chiral symmetry $\sigma_z$ and the intravalley inversion symmetry $\mI$. As defined in Table~\ref{emergentsym}, they respectively imply a particle-hole symmetry and an inversion symmetry to the spectrum and eigen-wavefunctions. In the definition of $\mI$, $\mK$ is the complex conjugation operator and $\mP = \text{diag}\left[\tau_y,-\tau_y,...,(-)^{n-1}\tau_y\right]$, where $\tau_y$ acts on the $i$th bottom and top layers. Throughout this work, we use Pauli matrix $\bm\sigma$ for sublattice and $\bm\tau$ for layers. Note that the intravalley inversion $\mI$ reduces to the known form for cTBG~\cite{JieWang_NodalStructure} when the $\mC_2\mT$ symmetry is restored~\footnote{Intravalley inversion in Ref.~\cite{JieWang_NodalStructure} is defined as $\mI^{\prime}\equiv\text{diag}\left(\mP,-\mP\right)$ which differs from $\mI$ by $\sigma_x\mK$, a combination of $\mC_2\mT$ and $\bm r\rightarrow-\bm r$. $\mC_2\mT$ invariance of $H_1$ implies $\mI^{\prime}H_1(\bm r)\mI^{\prime\dag}=H_1(-\bm r)$ which is consistent with Ref.~\cite{JieWang_NodalStructure}.}. Ignoring the negligible small twist angle effect, $\mI$ is identical to the approximate unitary particle-hole symmetry~\cite{Zhida_PRL19,Bernevig_tbg1,song2020tbg,bernevig2020tbg,lian2020tbg,bernevig2020tbg_5,xie2020tbg}.

\begin{table}
    \def\arraystretch{1.2}
    \begin{tabular}{c|c}
    \hline\hline
        \noalign{\vskip 0.5mm}
        \quad Chiral Symmetry \quad & \quad Intravalley Inversion Symmetry \quad \\
        \noalign{\vskip 0.5mm}
        \hline
        \noalign{\vskip 1.5mm}
        $\sigma_z=\left(\begin{matrix}1&\\&-1\end{matrix}\right)$  & $\mI = \left(\begin{matrix}\mP&\\&-\mP\end{matrix}\right)\sigma_x\mK$\\
        \noalign{\vskip 1.5mm}
        \hline
        \noalign{\vskip 1mm}
        $\{H,\sigma_z\}=0$ & $[H,\mI]=0$\\
        \noalign{\vskip 1mm}
        \hline\hline
    \end{tabular}\caption{Exact emergent symmetries of the model, which include the chiral symmetry $\sigma_z$ and the intra-valley inversion symmetry $\mI$. The combination of both symmetries implies that the spectrum is not only particle-hole symmetric but also $\bm k$ to $-\bm k$ symmetric within the same valley. The matrices above are written in the sublattice basis $\Phi$ and $\Xi$.}
    \label{emergentsym}
\end{table}

\emph{Twisted bilayer graphene.---}We now proceed to demonstrate the existence of ideal flatbands in our model. For the simplest case $n=1$ cTBG, Refs.~\cite{Grisha_TBG,Grisha_TBG2,JieWang_NodalStructure,becker2020mathematics,Becker_PRB21,RafeiRen_TBG,popov2020hidden,Gerardo21PRB} show that it is an exactly solvable model exhibiting $|\mC|=1$ dispersionless bands at charge neutrality with ideal band geometry at magic angles. Its wavefunction has an exact representation~\cite{JieWang_NodalStructure} (up to normalization) in terms of the LLL wavefunction $\Phi^{\rm LLL}_{\bm k}(\bm r)$~\cite{Jie_MonteCarlo,haldanemodularinv,scottjiehaldane,Jie_Dirac}:
\begin{equation}
\Phi_{1} = \left(\begin{matrix}\phi_1\\\phi'_1\end{matrix}\right) = \left(\begin{matrix}i\mG(\bm r)\\\eta\mG(-\bm r)\end{matrix}\right)\Phi^{\rm LLL}_{\bm k}(\bm r),\label{ctbgwf}
\end{equation}
where $\eta=\pm1$ is the intravalley inversion eigenvalue and the ${\bm k}$-independent $\mG(\bm r)$ can be interpreted as a quantum Hall wave function in a magnetic field oppositely directed to that of $\Phi^{\rm LLL}_{\bm k}(\bm r)$~\cite{JieWang_NodalStructure}. This connection to the LLL wavefunction implies that its cell periodic wavefunction $e^{-i\bm k\cdot\bm r}\Phi_{1,\bm k}$ is holomorphic in $k=(k_x+ik_y)/\sqrt2$ ignoring the normalization factor~\footnote{Following Refs.~\cite{haldanemodularinv,Jie_MonteCarlo,scottjiehaldane,Jie_Dirac}, the LLL wavefunction $\Phi^{\rm LLL}_{\bm k}(\bm r)$ can be expressed in terms of the modified Weierstrass sigma function $\sigma(z)$ as $e^{ik^*z}\sigma(z+ik)e^{-\frac{1}{2}(|z|^2+|k|^2)}$. Its ``cell-periodic'' part $u^{\rm LLL}_{\bm k}(\bm r)\equiv e^{-i\bm k\cdot\bm r}\Phi^{\rm LLL}_{\bm k}(\bm r)$ is holomorphic in $k$ ignoring the normalization factor.}. For any Bloch wavefunction satisfying this property, Eq.~(\ref{idealcond}) is automatically satisfied~\cite{PhysRevB.90.165139,Martin_PositionMomentumDuality,Grisha_TBG2,JieWang_exactlldescription}. The unit Chern number and ideal band geometry thereby make cTBG an exact $\bm k$-space dual of the LLL with nontrivial curvature~\cite{JieWang_exactlldescription}.

\emph{Twisted double bilayer graphene.---}We now discuss the first nontrivial case, i.e., $n=2$ cTDBG. It has been noticed that cTDBG has two exactly flat bands at charge neutrality~\cite{TDBG_nanoletter}. Despite of this observation, the wavefunction, topology and geometry of these flatbands were ignored before, which we will analyze in detail below. Since the two flatbands are sublattice polarized and related by $\mI$, without loss of generality we focus on the sublattice-$A$ flatband wavefunction $\Phi_2$ which is the zero mode of $\mD^{\dag}_2$. We denote $\Phi_2$ as $\left(\tilde\Phi^T_1,\phi_2,\phi_3\right)^T$ where $\tilde\Phi_1=(\tilde\phi_1,\tilde\phi'_1)^T$ is a two-component layer spinor. Component-wisely, the zero mode equation $\mD^{\dag}_2\Phi_2=0$ becomes
\begin{eqnarray}
\mD^{\dag}_1\tilde\Phi_1+t_1\left(0,~\phi_3\right)^T = 0,\quad-i\bar\partial\phi_3 &=& 0,\label{H2_sol0}\\
-i\bar\partial\phi_2 + t_1\tilde\phi_1 &=& 0.\label{H2_sol}
\end{eqnarray}

The solutions of these equations are $\phi_3=0$~\footnote{$\phi_3$ cannot be a non-zero constant, which violates the Bloch transnational symmetry.} and $\tilde\Phi_1$ being annihilated by $\mathcal{D}_1^{\dag}$. So $\tilde\Phi_1$ is identical to the cTBG wavefunction up to a normalization factor: $\tilde\Phi_1=N_{\bm k}\Phi_1$. For $N_{\bm k}\neq0$ one can rescale $\Phi_2$, so we replace $\tilde\Phi_1$ by $\Phi_1$ in below. As only $\mD_1^\dagger$ depends on the twist angle, the magic angles of cTDBG and cTBG are identical, at which the bands at charge neutrality are exactly flat.

The only nontrivial zero mode equation for cTDBG is Eq.~(\ref{H2_sol}) which governs the essential properties of band topology, band geometry and interacting physics through $\phi_2$. To prove the ideal band geometry of the magic angle cTDBG, we merely need to show the cell-periodic part of $\phi_2$ ($u_{1,2}\equiv e^{-i\bm k\cdot\bm r}\phi_{1,2}$) is holomorphic in $k$ up to a normalization, since $\Phi_1$, as the zero mode of cTBG, is already proved to satisfy this condition~\cite{Grisha_TBG2}. The key observation is that Eq.~(\ref{H2_sol}) only has anti-holomorphic derivative $\bar\partial$, thereby the differential equation for $u_2$, $\left(\bar\partial + ik\right)u_{2} = -it_1u_{1}$, depends only on $k$ but not on $\bar k$. Then $\bar\partial_k u_2 = 0$ follows immediately from the fact that $\bar\partial_k u_1=0$. At momentum points where $N_{\bm k}=0$, the zero mode equation $(\bar\partial+ik)u_2=0$ also immediately implies the $\bm k-$space holomorphic property of $u_2$ and thus the ideal band geometry of $\Phi_2$.

\begin{figure}
    \centering
    \includegraphics[width=\linewidth]{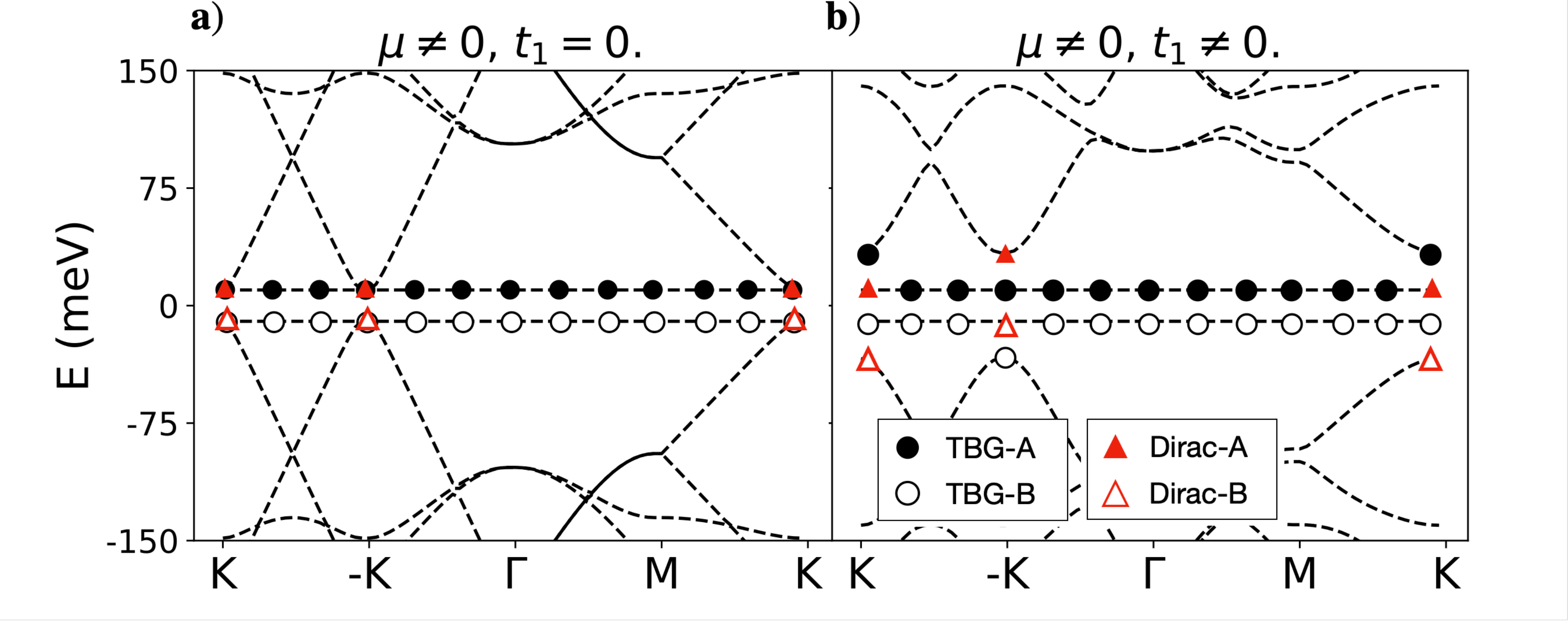}
    \caption{High-$\mC$ bands generated by the ``wavefunction exchange'' mechanism. In (a) and (b), we show the sublattice polarization properties of the cTBG (black circles) and Dirac wavefunctions (red triangles) before and after turning on an infinitesimal interlayer coupling $t_1$, respectively, where the solid/empty markers represent sublattice-$A$/$B$ polarization, respectively. The Dirac wavefunction interchanges with the cTBG wavefunction at $\pm\bm K$, which punctures a zero to cTBG wavefunction and increases the Chern number by one.}
    \label{fig:perturbation}
\end{figure}

Next we discuss band topology. While it is known that the Bernal-stacking structure can support high Chern number~\cite{Senthil_NearlyFlatBand,TDBG_nanoletter,XiDai_OM}, here we provide a proof which highlights the analytical structure of the cTDBG flatband wavefunction. For convenience, in the following we assume a small hexagonal-boron-nitride potential $\mu>0$ to split the degeneracy of the two cTDBG flatbands and meanwhile preserve their sublattice polarization.

We start by considering the limit of zero interlayer coupling $t_1=0$. In this case, in the low-energy regime there are two exactly flat bands $(\phi_{\rm cTBG}, \chi_{\rm cTBG})$ originating from the inner cTBG layers and two Dirac bands $(\phi_{\rm D},\chi_{\rm D})$ from the outermost layers. The cTBG and the Dirac bands are degenerate at the Dirac points $\pm\bm K$, as shown in Fig.~\ref{fig:perturbation}(a). In the following, we focus on the Dirac point $\bm K$ to examine the gap opening mechanism as the physics at $-\bm K$ is simply implied by the intravalley inversion. The $\bm K$ point wavefunctions $(\phi_{\rm cTBG}, \phi_{\rm D})$ at energy $\mu$ are sublattice-$A$ polarized and $(\chi_{\rm cTBG},\chi_{\rm D})$ at energy $-\mu$ are sublattice-$B$ polarized [Fig.~\ref{fig:perturbation}(a)]. We further note that $(\phi_{\rm D},\chi_{\rm D})$ are also polarized in the bottom layer. Under this scenario, in the ``$(\rm bottom, \rm top)$'' layer basis we have
\begin{eqnarray}
\phi_{\rm cTBG} &=& (\phi_1,\phi'_1)^T,\quad \phi_{\rm D} = (1,0)^T,\nonumber\\
\chi_{\rm cTBG} &=& (\chi_1,\chi'_1)^T,\quad \chi_{\rm D} = (1,0)^T.\label{pert_wf}
\end{eqnarray}

We then turn on an infinitesimal $t_1$ and use the perturbation theory to study the change of band structure and wavefunctions. As the $t_1$ terms couple adjacent layers of opposite sublattices, the perturbation matrix elements within the four low-energy bands are
\begin{equation}
    \langle\phi_{\rm cTBG}|T_+|\chi_{\rm D}\rangle \neq 0,\quad\langle\chi_{\rm cTBG}|T_-|\phi_{\rm D}\rangle = 0,\label{pert_eq}
\end{equation}
where details of Eq.~(\ref{pert_wf}) and Eq.~(\ref{pert_eq}) can be found in the SM.

Equation~(\ref{pert_eq}) implies that $\chi_{\rm cTBG}$ and $\phi_{\rm D}$ are unperturbed at $\bm K$, but $\phi_{\rm cTBG}$ and $\chi_{\rm D}$ start to repel each other immediately after turning on $t_1$. The net result is that a band gap is opened and the cTBG and Dirac bands at positive energy are effectively ``exchanged'' at $\bm K$ [Fig.~\ref{fig:perturbation}(b)], leaving $\Phi_{2,\bm K}$ to be $(0,0,1,0)^T$. We find that $\Phi_{2,\bm K}$ remains $(0,0,1,0)^T$ for arbitrary $t_1\neq0$ because the flatband energy stays at $\mu$ independent of $t_1$. On the other hand, Eq.~(\ref{H2_sol0}) dictates that first two components of $\Phi_2$ are identical to the cTBG wavefunction up to a normalization factor $N_{\bm k}$. Thus our analysis shows $N_{\bm k}$ must be zero at $\bm K$; how fast $N_{\bm k}$ decays to zero when $\bm k$ approaching $\bm K$ is determined by $|t_1|$.

This ``wavefunction exchange'' increases the flatband Chern number by one. The Chern number measures the discontinuity of the Bloch wavefunction which resides either at the boundary or in the bulk of the Brillouin zone~\cite{Thouless_1984,TKNN}. Since $\mC$ is an invariant, it is sufficient to work with an infinitesimal $|t_1|$. In this case, the cTDBG wavefunction is identical to cTBG wavefunction except near the Dirac points. One can choose the Brillouin zone boundary to avoid the Dirac points such that the boundary contribution to $\mC$ is determined by cTBG wavefunction which equals to one. The vanishing of $N_{\bm K}$ is equivalent as stating a pole singularity of the Dirac component $\phi_{\rm D}$ at $\bm K$, which increases the Chern number by one following Refs.~\cite{Martin_PositionMomentumDuality,Lee_engineering}. We therefore proved the cTDBG flatband has Chern number two.

\emph{Hierarchy scheme.---}The discussion of cTDBG ($n=2$) can be straightforwardly generalized to arbitrary $n$. Given the zero mode wavefunction $\Phi_{n-1}$ of $H_{n-1}$, the zero mode of $H_n$ must exist at the same magic angle, whose ansatz can be written as $\Phi^T_n = \left(\begin{matrix}\Phi^T_{n-1}, & \phi_n, & 0\end{matrix}\right)$ and the zero-mode equation generalizing Eq.~(\ref{H2_sol}) is
\begin{equation}
-i\bar\partial\phi_n + t_{n-1} \phi_{n-1} = 0.\label{zeromodeeqnforn}
\end{equation}

Since Eq.~(\ref{zeromodeeqnforn}) only has anti-holomorphic derivatives, the cell-periodic part of $\phi_n$ is a holomorphic function of $k$ as that of $\phi_{n-1}$ is. We therefore prove the ideal band geometry of $\Phi_{n}$ from the hierarchy construction. The band topology can also be analyzed by the same method. Starting with $t_{n-1}=0$, $t_{i=1,...,n-2}\neq0$, the $H_n$ at magic angle consists of two sublattice polarized flatbands originating from $\Phi_{n-1}$ which are degenerate with the two outermost freestanding Dirac bands at Dirac points. Finite but infinitesimal $|t_{n-1}|$ splits the degeneracy and ``exchanges'' the $\Phi_{n-1}$ with the Dirac band leaving $\Phi_{n,\bm K}$ to be $(0,...,0,1,0)^T$. This does not alter the boundary contribution to $\mC$ but generates an unavoidable bulk pole singularity and increases $\mC$ by one. We therefore prove the Chern number of our flatband equals to the number of layers and all the flatbands have ideal band geometry satisfying Eq.~(\ref{idealcond}). These results do not require infinitesimal $t_{i=1,...,n}$, because Chern number is a topological invariant and the ideal geometry follows directly from the holomorphic property of the zero-mode equations.

\begin{figure}
\centering
\includegraphics[width=\linewidth]{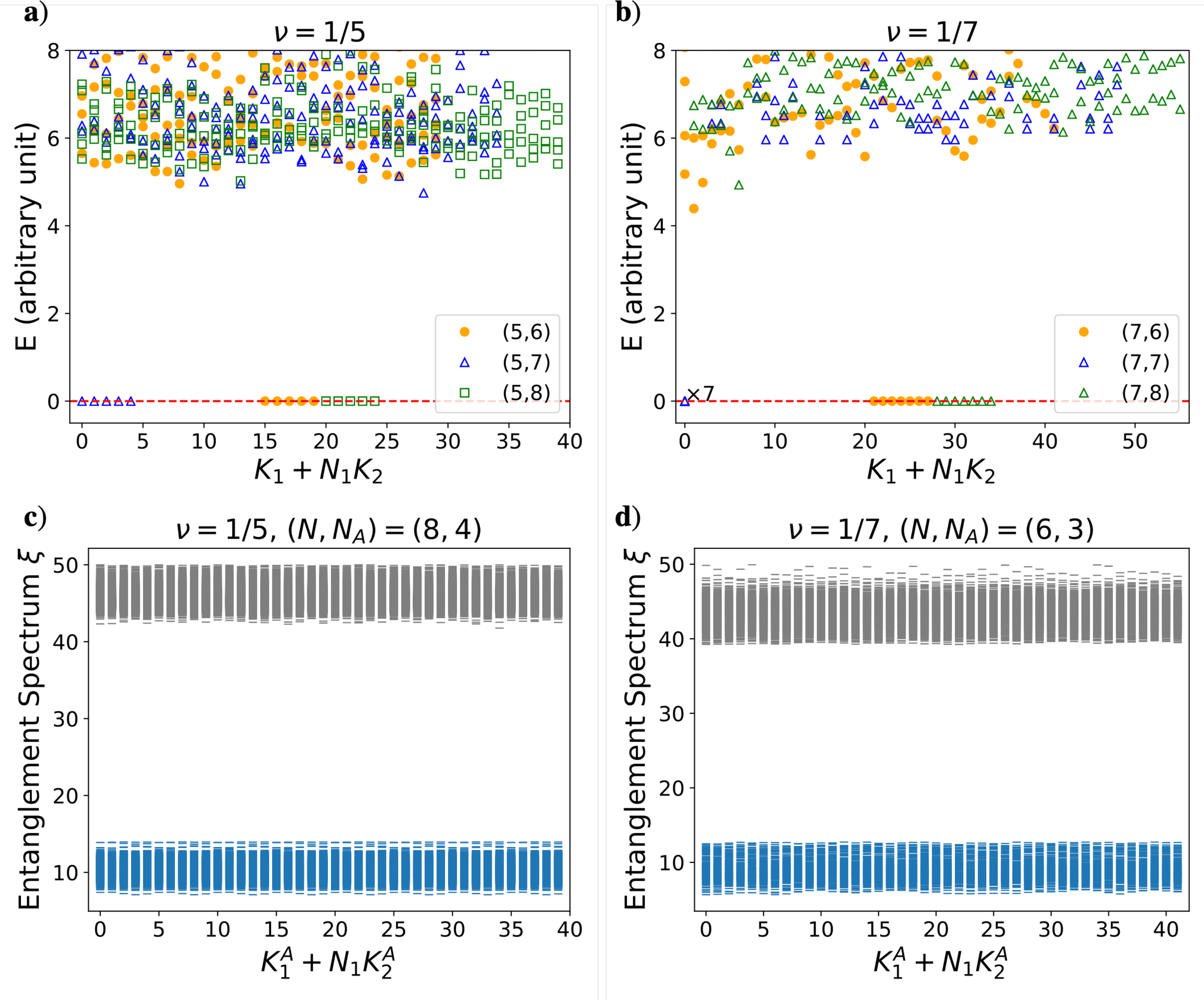}
\caption{Energy spectra and particle-cut entanglement spectra (PES) demonstrating exact model FCIs. (a) and (b): Energy spectra of the repulsive interaction $H_{\rm int}=-\sum_{i<j}\delta^{\prime\prime}(\bm r_i-\bm r_j)$ in the $\mC=n$ ideal flatband at filling fraction $\nu=1/(2n+1)$, where (a) is for $n=2$ (cTDBG) and (b) is for $n=3$ (chiral twisted double tri-layer graphene, cTDTG). An $(2n+1)$-fold exactly degenerate ground states at zero energy are clearly observed. Lattice sizes $(N_1, N_2)=(2n+1,N)$ are given in the legends. The red dashed lines mark the zero energy and are used to guide the eyes. (c) and (d): PES for $N=8$ and $N=6$ particles. The grey levels above $\xi_c=|\ln(2^{-53})|\approx36.7$ are machine noises. The number of low-energy PES levels is $17710$ and $3248$ in (c) and (d) respectively, agreeing with the FCI quasihole counting~\cite{Zhao_FCI_highC,Sterdyniak13}.}\label{fig:edfci}
\end{figure}

\emph{Exact fractional Chern insulators.---}We now examine the interacting physics in the ideal flatband of our model. As the pertinent band is exactly flat, we drop the kinetic energy and project the interaction into the ideal flatband. The band filling factor $\nu$ is defined as $N/(N_1 N_2)$ for $N$ electrons and $N_1,N_2$ unit cells in the two primitive directions of the \mr pattern. As the many-body Hamiltonian preserves the total momentum, each eigenstate can be labeled by its total momentum $(K_1,K_2)$. In TBG, it has been numerically demonstrated that the $\mC=1$ flatband at the charge neutrality can host the lattice Laughlin FCIs at $\nu=1/3$ \cite{ZhaoLiu_TBG,Cecile_TBG_Flatband,Cecile_PRL20}. In particular, the model $\nu=1/3$ Laughlin state was found to be the exact zero-energy ground state at the chiral limit for the short-ranged two-body repulsive interaction $H_{\rm int}=-\sum_{i<j}\delta^{\prime\prime}(\bm r_i-\bm r_j)$~\cite{Haldane_hierarchy}.

In high-$\mC$ Bloch bands, robust FCIs were reported across various models~\cite{Zhao_FCI_highC,Sterdyniak13,Yangle_Modelwf,Moller_FCI_highC,YinghaiWu_HighC,ModelFCI_Zhao,Bart_HighC,Bart_stability,Bart_FQHtransition}. Remarkably, in our ideal flatbands, we observe \emph{exact} $(2n+1)$-fold degenerate zero-energy ground states for $H_{\rm int}$, separated by a finite energy gap to excitations [Figs.~\ref{fig:edfci}(a) and \ref{fig:edfci}(b) for $n=2$ and $n=3$]. Their particle-cut entanglement spectra (PES)~\cite{PhysRevX.1.021014}, defined as the entanglement between subsystems of $N_A$ and $N-N_A$ particles, are displayed in Figs.~\ref{fig:edfci} (c) and \ref{fig:edfci}(d). The counting of low PES levels agrees with the expectation from FCI quasihole excitations. The high PES levels appear only above the machine error cut-off $\xi_c\approx36.7$, strongly suggesting an infinite PES gap and the exact zero modes are model FCIs. See the SM for studies away from the chiral limit.

\emph{Discussions.---}
There are a couple of open questions which deserve future studies. We noticed that the model FCIs are intrinsic to the outermost Dirac layer: further projecting $H_{\rm int}$ into the $\phi_n$ component of $\Phi_n$ changes the energies of excited states but leaves the exact degenerate zero-energy ground states and the PES unaffected. This means $\phi_n$ alone could exhibit a ``color-entangled'' feature~\cite{Yangle_Modelwf} which remains challenging to uncover analytically from the zero mode equation Eq.~(\ref{zeromodeeqnforn}). Furthermore, a thorough understanding of the origin of the exact model FCIs is still lacking. Exact model FCIs were also reported in the numerical studies of onsite interacting bosons in the Kapit-Mueller model~\cite{Kapit_Mueller,Dong_Mueller_20} and its variations~\cite{ModelFCI_Zhao}. Considering the band geometry of the Kapit-Mueller model is also ideal~\cite{Emil_constantBerry}, we anticipate the ideal geometry is the fundamental origin of the frustration free nature of these lattice-specific interacting Hamiltonians. Studying the projected density algebra is an interesting future direction~\cite{gmpl,gmpb,Haldanegeometry,Cecil_singlemodeapp,Nicolas_13PRB}.

\begin{acknowledgments}
J.W. is grateful to Bartholomew Andrews, Semyon Klevtsov, Nicolas Regnault, and Ya-Hui Zhang for useful discussions. We acknowledge Jennifer Cano, Andrew J. Millis, and Bo Yang for the collaboration on Ref.~(\onlinecite{JieWang_exactlldescription}). Z.L. acknowledges Ahmed Abouelkomsan and Emil J. Bergholtz for the collaboration on related topics. Z.L. is supported by the National Key Research and Development Program of China through Grant No. 2020YFA0309200. The Flatiron Institute is a division of the Simons Foundation. The authors are grateful to Lucy Reading-Ikkanda for creating the cover figure.

\emph{Note added:} after the completion of this work, Ref.~\cite{Eslam_highC_idealband} appeared, which overlaps with the results reported here.
\end{acknowledgments}

\bibliography{ref.bib}

\clearpage
\appendix
\onecolumngrid
\section*{------ APPENDIX ------}
We provide necessary details in this supplementary material. They include: (1) the unitary transformations that lead to our multilayered chiral model, (2) the algebra for the intravalley inversion symmetry, (3) numerical verification of ideal quantum geometries in high-$\mC$ bands, (4) details of the wavefunction exchange mechanism, (5) numerical details for interacting physics and (6) evolution of band geometry and interacting effects away from the chiral limit.

\section{Chiral Hamiltonian}\label{app:chiralH}
In this section, we discuss the details of the cTDBG Hamiltonian. The generalization to arbitrary $n$ is straightforward. In the basis (Dirac bottom, TBG bottom, TBG top, Dirac top), the chiral Hamiltonian is given by,
\begin{equation}
h_0(\bm r) = \left(\begin{matrix}\bm{\sigma}_{-\theta/2}(-i\bm\nabla-\bm K_+^b)&T_0&0&0\\T^{\dag}_0&\bm{\sigma}_{-\theta/2}(-i\bm\nabla-\bm K_+^b)&T_{\theta}(\bm r)&0\\0&T^{\dag}_{\theta}(\bm r)&\bm{\sigma}_{+\theta/2}(-i\bm\nabla-\bm K_+^t)&T_0\\0&0&T^{\dag}_0&\bm{\sigma}_{+\theta/2}(-i\bm\nabla-\bm K_+^t)\end{matrix}\right),
\end{equation}
where the parameters are chosen as those in the standard Bistritzer-MacDonald model. $\bm K_{b,t}$ are respectively the \mr Dirac point contributed from the bottom and top layer. Concrete values of the parameters can be found for instance in the appendix of Refs.~\cite{ZhaoLiu_TBG,ZhaoLiu_TDBG}. In the above Hamiltonian,
\begin{equation}
T_0 = \left(\begin{matrix} 0&0\\t_1&0 \end{matrix}\right),\quad T_{\theta}^{\dag}(\bm r) = \sum_{j=0}^2T_{j+1}e^{-i(\bm q_0-\bm q_j)\cdot\bm r},\quad T_{j+1} = \omega_1e^{i(2\pi/3)j\sigma_z}\sigma_xe^{-i(2\pi/3)j\sigma_z}.
\end{equation}

Following Ref.~\cite{Grisha_TBG}, we perform unitary transformations to remove the $\bm K_{b/t}$ and $\theta$ dependence:
\begin{eqnarray}
h_0(\bm r) &=& \mathcal{M}h_1(\bm r)\mathcal{M}^{\dag},\quad\mathcal{M} = \mathcal{M}_T\mathcal{M}_{\theta}\mathcal{M}_D,\label{translationgauge}\\
\mathcal{M}_T &=& \text{diag}\left(e^{i\bm K^b_+\cdot\bm r},e^{i\bm K^b_+\cdot\bm r},e^{i\bm K^t_+\cdot\bm r},e^{i\bm K^t_+\cdot\bm r}\right),\nonumber\\
\mathcal{M}_{\theta} &=& \text{diag}\left(e^{i\frac{\theta}{4}\sigma_z},e^{i\frac{\theta}{4}\sigma_z},e^{-i\frac{\theta}{4}\sigma_z},e^{-i\frac{\theta}{4}\sigma_z}\right).\nonumber
\end{eqnarray}

The transformed Hamiltonian $h_1(\bm r)$ reads:
\begin{equation}
h_1(\bm r) = \left(\begin{matrix}-i\bm{\sigma}\cdot\bm\nabla&T_{0,\theta}&0&0\\T^{\dag}_{0,\theta}&-i\bm{\sigma}\cdot\bm\nabla&T_{\theta}(\bm r)&0\\0&T^{\dag}_{\theta}(\bm r)&-i\bm{\sigma}\cdot\bm\nabla&T_{0,\theta}\\0&0&T^{\dag}_{0,\theta}&-i\bm{\sigma}\cdot\bm\nabla\end{matrix}\right) = \left(\begin{matrix}
0&-i\partial&0&0&0&0&0&0\\
-i\bar{\partial}&0&t_{\theta}&0&0&0&0&0\\
0&t^*_{\theta}&0&-i\partial&0&U_{-\phi}&0&0\\
0&0&-i\bar{\partial}&0&U_{\phi}&0&0&0\\
0&0&0&U^*_{\phi}&0&-i\partial&0&0\\
0&0&U^*_{-\phi}&0&-i\bar{\partial}&0&t^*_{\theta}&0\\
0&0&0&0&0&t_{\theta}&0&-i\partial\\
0&0&0&0&0&0&-i\bar{\partial}&0
\end{matrix}\right),
\end{equation}
where $T_{0,\theta}=\left(\begin{matrix}0&0\\t_{\theta}&0\end{matrix}\right)$ and $t_{\theta}=\exp(i\theta/2)$. We next shuffle into the sublattice basis, and the Hamiltonian is brounght into the off-diagonal form:
\begin{equation}
h_3(\bm r) = \left(\begin{matrix}&\mD_{2,\theta}\\\mD_{2,\theta}^{\dag}&\end{matrix}\right),\quad\mathcal{D}^{\dag}_{2,\theta} = \left(\begin{matrix}-i\bar{\partial}&t_{\theta}&0&0\\0&-i\bar{\partial}&U_{\phi}&0\\0&U^*_{-\phi}&-i\bar{\partial}&t^*_{\theta}\\0&0&0&-i\bar{\partial}\end{matrix}\right),\quad\mathcal{D}_{2,\theta} = \left(\begin{matrix}-i\partial&0&0&0\\t^*_{\theta}&-i\partial&U_{-\phi}&0\\0&U^*_{\phi}&-i\partial&0\\0&0&t_{\theta}&-i\partial\end{matrix}\right).
\end{equation}

The $\theta$ dependence in $t_{\theta}$ can be further eliminated by a unitary rotation in the Dirac layers:
\begin{equation}
\mathcal{D}^{\dag}_2 \equiv \left(\begin{matrix}e^{-i\theta/2}&0&0&0\\0&1&0&0\\0&0&1&0\\0&0&0&e^{-i\theta/2}\end{matrix}\right)\mathcal{D}^{\dag}_{2,\theta}\left(\begin{matrix}e^{i\theta/2}&0&0&0\\0&1&0&0\\0&0&1&0\\0&0&0&e^{i\theta/2}\end{matrix}\right),
\end{equation}
after which we arrive at the chiral-TDBG Hamiltonian used in the main text:
\begin{equation}
H_{n=2}(\bm r) = \left(\begin{matrix}&\mD_2\\\mD^{\dag}_2&\end{matrix}\right),\quad\mathcal{D}^{\dag}_2 = \left(\begin{matrix}-i\bar{\partial}&t_1&0&0\\0&-i\bar{\partial}&U_{\phi}&0\\0&U^*_{-\phi}&-i\bar{\partial}&t_1\\0&0&0&-i\bar{\partial}\end{matrix}\right),\quad\mathcal{D}_2 = \left(\begin{matrix}-i\partial&0&0&0\\t_1&-i\partial&U_{-\phi}&0\\0&U^*_{\phi}&-i\partial&0\\0&0&t_1&-i\partial\end{matrix}\right).\nonumber
\end{equation}

The multi-layer generalization of the above unitary transformations is straightforward, leading to the chiral model Hamiltonian,
\begin{equation}
H_n(\bm r) = \left(\begin{matrix} & \mD_n \\ \mD_n^{\dag} &\end{matrix}\right),\quad\quad\mD_{n} = \left(\begin{matrix}\mD_1 & t_1T_+ & \\ t_1T_- & h_D & t_2T_+ & \\ & t_2T_- & h_D & \ddots \\ && \ddots && \\ &&&& t_{n-1}T_+ \\ &&&t_{n-1}T_- & h_D\end{matrix}\right).\label{app_model}
\end{equation}

\section{Intravalley Inversion}
Here we show the details of $[\mI, H_n]=0$ where $\mI$ is the intravalley inversion operation:
\begin{equation}
\mI = \left(\begin{matrix}&\mP\\-\mP&\end{matrix}\right)\mK,\quad\mP=\text{diag}\left(\tau_y,-\tau_y,...,(-)^{n-1}\tau_y\right).
\end{equation}

We first note that:
\begin{equation}
\mP\mathcal{D}_n(\bm r)\mP = -\left(\mathcal{D}_n^{\dag}(\bm r)\right)^*,\quad \mP\mathcal{D}_n^{\dag}(\bm r)\mP = -\mathcal{D}_n^*(\bm r).\label{PDP}
\end{equation}

The derivation of Eq.~(\ref{PDP}) can be seen as below:
\begin{eqnarray}
\mP \mD_n\mP &=& \left(\begin{matrix}\tau_y\mD_1\tau_y\quad & -t_1\tau_yT_+\tau_y\quad & \\ -t_1\tau_yT_-\tau_y\quad & \tau_yh_D\tau_y\quad & -t_2\tau_yT_+\tau_y\quad & \\ & -t_2\tau_yT_-\tau_y\quad & \tau_yh_D\tau_y\quad & \ddots \\ && \ddots && \\ &&&& -t_{n-1}\tau_yT_+\tau_y\quad \\ &&& -t_{n-1}\tau_yT_-\tau_y\quad & \tau_yh_D\tau_y\quad\end{matrix}\right),\nonumber\\
&=& -\left(\begin{matrix}\mD^{\dag}_1 & t_1T_- & \\ t_1T_+ & h^{\dag}_D & t_2T_- & \\ & t_2T_+ & h^{\dag}_D & \ddots \\ && \ddots && \\ &&&& t_{n-1}T_- \\ &&&t_{n-1}T_+ & h^{\dag}_D \end{matrix}\right)^* = -[\mD^{\dag}_n]^*,
\end{eqnarray}
where in the last line, we used the identity \cite{JieWang_NodalStructure} following the $\mC_2\mT$ invariance of cTBG,
\begin{equation}
    \tau_y D_1(\bm r)\tau_y = -\mD_1(-\bm r) = -[\mD_1^{\dag}(\bm r)]^* ,\quad \tau_y h_D(\bm r)\tau_y = -h_D(-\bm r) = -[h_D^{\dag}(\bm r)]^*.\label{identity}
\end{equation}

Therefore we can prove $[\mI,H_n]=0$ by using:
\begin{equation}
\left(\begin{matrix}&\mP\\-\mP&\end{matrix}\right)\left(\begin{matrix}&\mD_n\\\mD_n^{\dag}&\end{matrix}\right)\left(\begin{matrix}&\mP\\-\mP&\end{matrix}\right) = \left(\begin{matrix}&-\mP\mD_n^{\dag}\mP\\-\mP\mD_n\mP&\end{matrix}\right) = \left(\begin{matrix}&\mD_n\\\mD_n^{\dag}&\end{matrix}\right)^*.
\end{equation}

\section{Numerical Calculation of Band Geometries}
In the main text, we have proved the band geometry of our model is ideal. This means that Berry curvature is positive and proportional to the Fubini-Study metric $\Omega_{\bm k} = \omega_{ab}g^{ab}_{\bm k}$. The right hand side defines the trace of the Fubini-Study metric.

The Berry curvature $\Omega_{\bm k}$ and the Fubini-Study metric $g^{ab}_{\bm k}$ are defined respectively as the imaginary and real part of the quantum geometric tensor,
\begin{equation}
\mathcal{Q}^{ab}_{\bm k} \equiv \langle D^a_{\bm k}u_{\bm k}|D^b_{\bm k}u_{\bm k}\rangle = g^{ab}_{\bm k} + \frac{i\epsilon^{ab}}{2}\Omega_{\bm k},
\end{equation}
where $|u_{\bm k}\rangle$ is the cell-periodic part of the Bloch wavefunction, $D^{a}_{\bm k}=\partial^a_{\bm k}-iA_{\bm k}^a$ is the covariant derivative with respect to the Berry connection $A^a_{\bm k} = -i\langle u_{\bm k}|\partial^a_{\bm k}u_{\bm k}\rangle$, and $\epsilon^{xy}=-\epsilon^{yx}=1$ is the 2D anti-symmetric tensor.

\begin{figure}[ht]
\centering
\includegraphics[width=\linewidth]{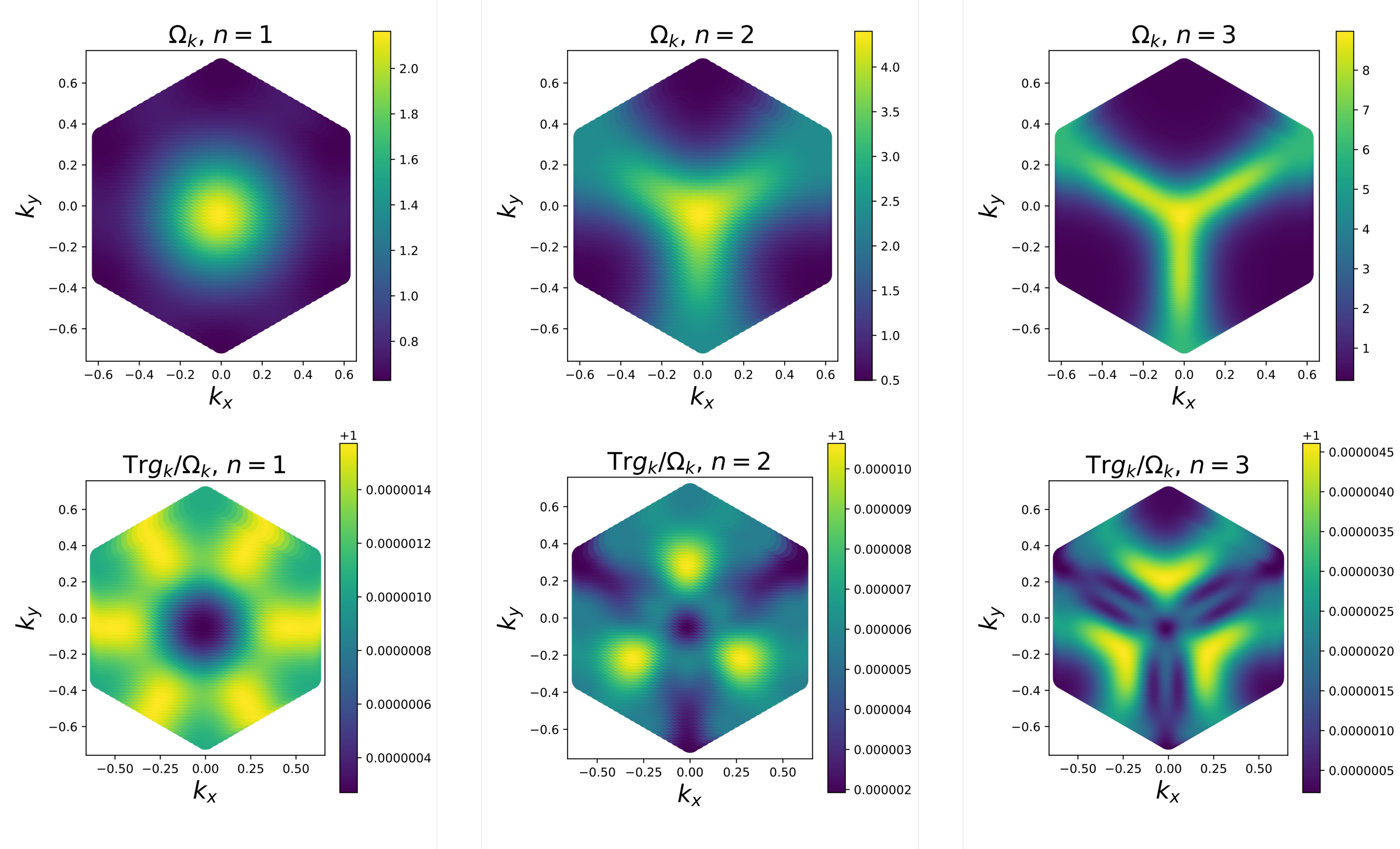}
\caption{The numerical data of Berry curvature (first row) and its ratio to the Fubini-Study metric $\text{Tr}g_{\bm k}/\Omega_{\bm k}=1$ (second row), in the flatband of cTBG (first column), cTDBG (second column) and cTDTG (third column). The trace is defined as $\text{Tr}g_{\bm k} = \omega_{ab}g^{ab}_{\bm k}$ where $\omega_{ab}$ in our case is $\delta_{ab}$. The numerical result verifies the ideal flatband condition that the Berry curvature is non-vanishing, and proportional to the trace of the Fubini-Study metric. Small deviations to $1$ of the second row are numerical artifacts due to finite momentum space grid in numerical calculations which goes to zero in the thermodynamic limit.}\label{sup:bandgeo}
\end{figure}

In Fig.~\ref{sup:bandgeo} we numerically verify the ideal band geometry for flatbands in our model for $n=1,2,3$ which corresponds to cTBG, cTDBG and chiral twisted double tri-layer graphene (cTDTG), respectively. The numerical result verifies the ideal flatband condition that the Berry curvature is non-vanishing, and proportional to the trace of the Fubini-Study metric.

\section{Details of Perturbation Analysis and the Wavefunction Exchange Mechanism}\label{app:pert}
In this section, we work out the details of the basis wavefunctions used for the degenerate perturbation calculation discussed in the main text. For simplicity, we focus on $n=2$ where the basis wavefunctions includes two Dirac wavefunctions and two cTBG flatband wavefunctions at the \mr $\pm\bm K$ points. We start with considering interlayer coupling $t_1=0$. The Dirac Hamiltonian describing the outermost two layers written in the sublattice basis is given by:
\begin{equation}
H_D = \left(\begin{matrix}\mu\tau_0&-i\partial\tau_0\\-i\bar\partial\tau_0&-\mu\tau_0\end{matrix}\right),
\end{equation}
where $\bm\tau$ is the layer Pauli matrix (representing the bottom-most and top-most layers), $\mu$ is the hexagonal-boron-nitride potential. At \mr Dirac point $\pm\bm K$, its eigenstate wavefunction is a tensor product,
\begin{equation}
    \Psi_{D} = \psi^{\sigma}\otimes\psi^{\tau},
\end{equation}
where $\psi^{\sigma}$ is a sublattice spinor and $\psi^{\tau}$ is a layer spinor.

Note that because of the unitary transformation Eq.~(\ref{translationgauge}), the action of translations in the chiral basis is:
\begin{equation}
\mathcal{V}_{1,2}\psi_{\bm k}(\bm r) = e^{i\tau_z\bm K\cdot\bm a_{1,2}}\psi_{\bm k}(\bm r+\bm a_{1,2}) = e^{i(\bm k+\tau_z\bm K)\cdot\bm a_{1,2}}\psi_{\bm k}(\bm r),\label{moiretranslation}
\end{equation}
where $\tau_z=+1$ for the bottom layers, and $\tau_z=-1$ for the top $n$ layers. Therefore the $\bm k=0$ point corresponds to the \mr $\bm K$ point for the bottom $n$ layers, and to the $-\bm K$ point for the top layers.

For this reason, $\psi^{\tau}$ has a simple form at \mr Dirac points $\bm k=0$ [written in basis of the (bom, top) layer]:
\begin{equation}
    \psi^{\tau}_{\bm k} = \begin{cases} (1,0)^T,\quad\text{for $\bm k=\bm K$}\\ (0,1)^T,\quad\text{for $\bm k=-\bm K$} \end{cases}.
\end{equation}

Since $\psi^{\tau}$ is a simple layer spinor, we focus on the sublattice wavefunctions $\psi^{\sigma}$ only. When $\mu=0$, the eigen-energies and eigen-wavefunctions are:
\begin{eqnarray}
E_- &=& -|k|,\quad \psi^{\sigma}_{-,\bm k} = \left(-\frac{|k|}{k},+1\right)^T,\nonumber\\
E_+ &=& +|k|,\quad \psi^{\sigma}_{+,\bm k} = \left(+\frac{|k|}{k},+1\right)^T,\nonumber
\end{eqnarray}
which has a singularity at Dirac point $\bm k=0$. The singularity can be removed by setting $\mu\neq0$. With $k=(k_x+ik_y)/\sqrt2$ and $\bar k=(k_x-ik_y)/\sqrt2$, the eigen-energies and eigen-wavefunctions are:
\begin{eqnarray}
E_- &=& -\sqrt{|k|^2+\mu^2},\quad \psi^{\sigma}_{-,\bm k} = \left(\frac{\mu-\sqrt{|k|^2+\mu^2}}{k},1\right)^T,\nonumber\\
E_+ &=& +\sqrt{|k|^2+\mu^2},\quad \psi^{\sigma}_{+,\bm k} = \left(1,\frac{-\mu+\sqrt{|k|^2+\mu^2}}{\bar k}\right)^T.
\end{eqnarray}

Reducing to the Dirac point ${\bm k}=0$, we get:
\begin{eqnarray}
E_- &=& -|\mu|,\quad \psi^{\sigma}_{-,\bm k=\bm 0} = (0,1)^T,\nonumber\\
E_+ &=& +|\mu|,\quad \psi^{\sigma}_{+,\bm k=\bm 0} = (1,0)^T,\label{Diracwf}
\end{eqnarray}
from which we see the positive (negative) energy state is completely sublattice-$A$ ($B$) polarized.

Writing in the basis of (TBG-A, Dirac-A, TBG-B, Dirac-B), we get the \mr $\bm K$ point Dirac wavefunction as below:
\begin{equation}
\Psi^+_{\rm D} = \left(\begin{matrix}\bm 0&\phi^T_{\rm D}&\bm 0&\bm 0\end{matrix}\right)^T,\quad\Psi^-_{\rm D} = \left(\begin{matrix}\bm 0&\bm 0&\bm 0&\chi^T_{\rm D}\end{matrix}\right)^T,\label{diracwavefunction}
\end{equation}
where $\phi_{\rm D}=\chi_{\rm D}=\left(\begin{matrix}1, & 0\end{matrix}\right)^T$ is a layer spinor representing the bottom and top layer. The cTBG wavefunctions are sublattice polarized, so we can label them by:
\begin{equation}
\Psi^+_{\rm cTBG} = \left(\begin{matrix}\phi^T_{\rm cTBG}&\bm 0&\bm 0&\bm 0\end{matrix}\right)^T,\quad\Psi^-_{\rm cTBG} = \left(\begin{matrix}\bm 0&\bm 0&\chi^T_{\rm cTBG}&\bm 0\end{matrix}\right)^T.
\end{equation}

In the (TBG, Dirac) basis, the perturbation matrix $H_T$ is given by,
\begin{equation}
H_T = \left(\begin{matrix}&T\\T^{\dag}&\end{matrix}\right),\quad T=\left(\begin{matrix}&T_+\\T_-&\end{matrix}\right),
\end{equation}
where $T_{\pm}$ are given in the main text. The $H_T$ couples distinct sublattice, whose matrix elements in terms of $\Psi^{\pm}_{\rm D}$ and $\Psi^{\pm}_{\rm cTBG}$ are,
\begin{eqnarray}
&&\langle\Psi^+_{\rm cTBG}|H_T|\Psi^-_{\rm D}\rangle = \langle\phi_{\rm cTBG}|T_+|\chi_{\rm D}\rangle \neq 0,\label{sup_pert1}\\
&&\langle\Psi^-_{\rm cTBG}|H_T|\Psi^+_{\rm D}\rangle = \langle\chi_{\rm cTBG}|T_-|\phi_{\rm D}\rangle = 0,\label{sup_pert2}
\end{eqnarray}
and their Hermitian conjugates. Eq.~(\ref{sup_pert1}) and Eq.~(\ref{sup_pert2}) are the key results of this section, used in the main text to analyze the analytical property of the Bloch wavefunction when $t_1$ is nonzero.


\section{Interacting Hamiltonian and Exact Fractional Chern Insulators}
In momentum space, the many-body interaction is,
\begin{eqnarray}
    H &=& \sum_{{\bm q}} v_{\bm q} :\rho_{\bm q} \rho_{-{\bm q}}:,\label{sup_H}\\
    v_{\bm q} &=& \sum_{m}\tilde{v}_m L_m(\bm q^2\ell^2)\exp(-\bm q^2\ell^2/2),\label{sup_vq}
\end{eqnarray}
where $\rho_{\bm q}$ is the flatband projected density operator and $::$ is the normal order. The $v_{\bm q}$ can be expanded into Haldane's pseudopotentials according to Eq.~(\ref{sup_vq}) where $L_m$ is the Laguerre polynomial and Laguerre-Gaussian functions are complete orthogonal basis. Here $2\pi\ell^2$ is the area of the \mr unit cell. The $H_{\rm int}$ used in the main text corresponds to the model interaction with only $\tilde{v}_1\neq0$.

In this appendix, we comment on the effect of $\tilde{v}_0$. For single-layer fermionic (bosonic) systems, only odd (even) components matter. For multi-layered systems, both even and odd components influence the energy spectrum, as the inter-layer interaction is not constrained by the Pauli principle.

For layer-isotropic interactions, the energy spectrum depends on both $\tilde{v}_0$ and $\tilde{v}_1$, but the exact zero modes observed in Fig.~3 of the main text do not depend on the precise values of $\tilde{v}_{0,1}$. We want to emphasis these exact FCIs are intrinsic to the outermost Dirac layer, {\it i.e.}, the $\phi_n$ component. We have checked that further projecting $\rho_{\bm q}$ into $\phi_n$ only affects the excited energies but retains the $(2n+1)$-fold degenerate zero modes and gives identical ground-state PES. Furthermore, the energy spectrum of the $\phi_n$ layer projected interaction is independent on the $v_0$ pseudopotential, implying the zero-energy ground states are purely spanned in the basis of $\phi_n$, with a vanishing power of $2n+1$ when two electrons approach each other.

\section{Beyond the Chiral Limit}
In the main text, we discussed the band geometry and the interacting physics of the Bernal stacked twisted multi-layered graphene model in the chiral limit. In this section, we study the band geometry and the interacting physics away from the chiral limit. We take the twisted double bilayer graphene (TDBG) as an example, which consists of a twisted bilayer graphene as the innermost two layers, and two outer layers of graphene in a Bernal stacking configuration.

The parameters we used for the chiral limit are:
\begin{itemize}
    \item
    $w_1=0.11 \rm{eV}$: the inter-sublattice tunneling strength between the innermost two layers.
    \item
    $t_0=2.61 \rm{eV}$: the nearest-neighbor hopping strength in monolayer graphene, which sets the Fermi velocity of monolayer graphene.
    \item
    $t_1=0.361 \rm{eV}$: the interlayer tunneling strength between the dimer sites in the Bernal stacking configuration between the innermost two layers and the outer two layers.
    \item
    $\theta$: the twist angle, which is set to be the first magic angle, {\it i.e.}, the largest angle at which the dispersion at charge neutrality becomes exactly zero. It takes the same value as the magic angle of chiral twisted bilayer graphene.
    \item
    $M=0.02 \rm{eV}$: the mass term which we use to split the degeneracy of the two flat bands at the charge neutrality. Such a term can be induced by two hexagonal boron nitride layers capsulating the TDBG.
\end{itemize}

In the realistic TDBG, there are other important parameters, including:
\begin{itemize}
    \item
    $w_0$: the intra-sublattice tunneling between the innermost two layers. Typically we have $w_0\approx 0.7w_1$ due to the lattice relaxation effect.
    \item
    $t_3$: the trigonal warping strength in the Bernal stacking configuration between the innermost two layers and the outer two layers. 
    \item
    The particle-hole asymmetry terms~\cite{PhysRevB.89.035405,Lee:2019aa} in the Bernal stacking configuration. These terms are weaker than the trigonal warping.
    \item
    An externally applied vertical bias voltage.
\end{itemize}
All of these additional parameters are zero in the chiral limit. More details of the realistic TDBG model can be found in Refs.~\cite{Lee:2019aa,ZhaoLiu_TDBG}.

\begin{figure}
\centering
\includegraphics[width=\linewidth]{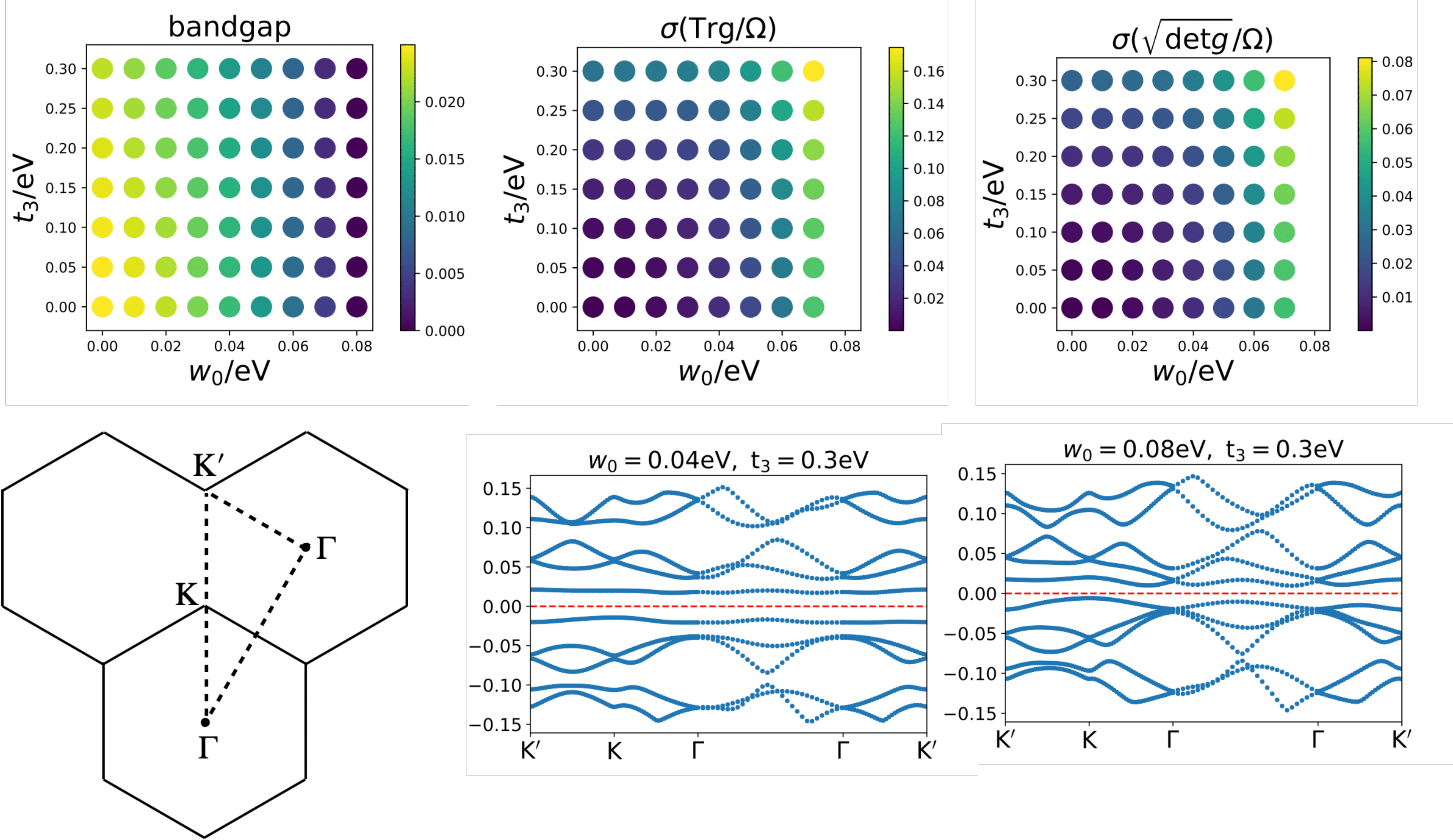}
\caption{First row: Indirect band gap, variation of $\text{Tr}g_{\bm k}/\Omega_{\bm k}$, and variation of $\sqrt{\det g_{\bm k}}/\Omega_{\bm k}$ of the first conduction band as a function of $(w_0,t_3)$. Second row: representative band structure at small $w_0$ with an isolated conduction band and at large $w_0$ without an isolated conduction band. Here we choose a specific path in the Brillouin zone. The red dotted line is the chemical potential $\mu=0$.}\label{sup:var_w0_t3}
\end{figure}

As shown above, there are a lot of parameters that can drive the system away from the chiral limit. It is complicated to thoroughly study their effects on the band geometry and the interacting physics. In this appendix, we focus on the effect of nonzero $(w_0,t_3)$. We first consider the band gap and band geometry near the charge neutrality. Since nonzero $(w_0,t_3)$ drives the system away from the chiral limit, the flatbands near the charge neutrality are no longer exactly flat and their band geometries are no longer ideal. Moreover, nonzero $(w_0,t_3)$ could induce gap closing in some parameter regimes. Focusing on the first conduction band above the charge neutrality, we plot the indirect band gap, $\sigma({\rm Tr} g/\Omega)$, and $\sigma(\sqrt{\det g}/\Omega)$ as functions of $(w_0,t_3)$ in the first row of Fig.~\ref{sup:var_w0_t3}, where $\sigma(O) \equiv \sqrt{\overline{O^2}-\overline{O}^2}$
quantifies the variation of the quantity $O$ in the Brillouin zone and $\overline{O}$ is the mean of $O$ in the Brillouin zone.  

We see the band gap closes when $w_0\geq0.08 \rm{eV}$, {\it i.e.}, the first conduction band above the charge neutrality is not isolated. In the second row of Fig.~\ref{sup:var_w0_t3}, typical band structures with and without an isolated first conduction band are given at $(w_0,t_3)=(0.04,0.3)\rm{eV}$ and $(w_0,t_3)=(0.08,0.3)\rm{eV}$, respectively. 

We calculate the band geometry of the first conduction band above the charge neutrality when it is isolated. In the chiral limit, both ${\rm Tr} g_{\bm k}/\Omega_{\bm k}$ and $\sqrt{\det g_{\bm k}}/\Omega_{\bm k}$ are constants independent of the Bloch momentum $\bm k$, so their variations in the Brillouin zone are strictly zero. Figs.~\ref{sup:var_w0_t3}(c) and ~\ref{sup:var_w0_t3}(d) show how their variations increase in the parameter space of $(w_0,t_3)$. Remarkably, over a wide range with $w_0\leq 0.05\rm{eV}$ and $t_3\leq 0.25\rm{eV}$, the band geometry is still close to the ideal case as both $\sigma({\rm Tr} g/\Omega)$ and $\sigma(\sqrt{\det g}/\Omega)$ are quite small. This implies exotic FCIs are stable in this parameter regime. We notice $w_0$ plays a more important role in driving the band geometry away from the ideal case. 

\begin{figure}
\centering
\includegraphics[width=\linewidth]{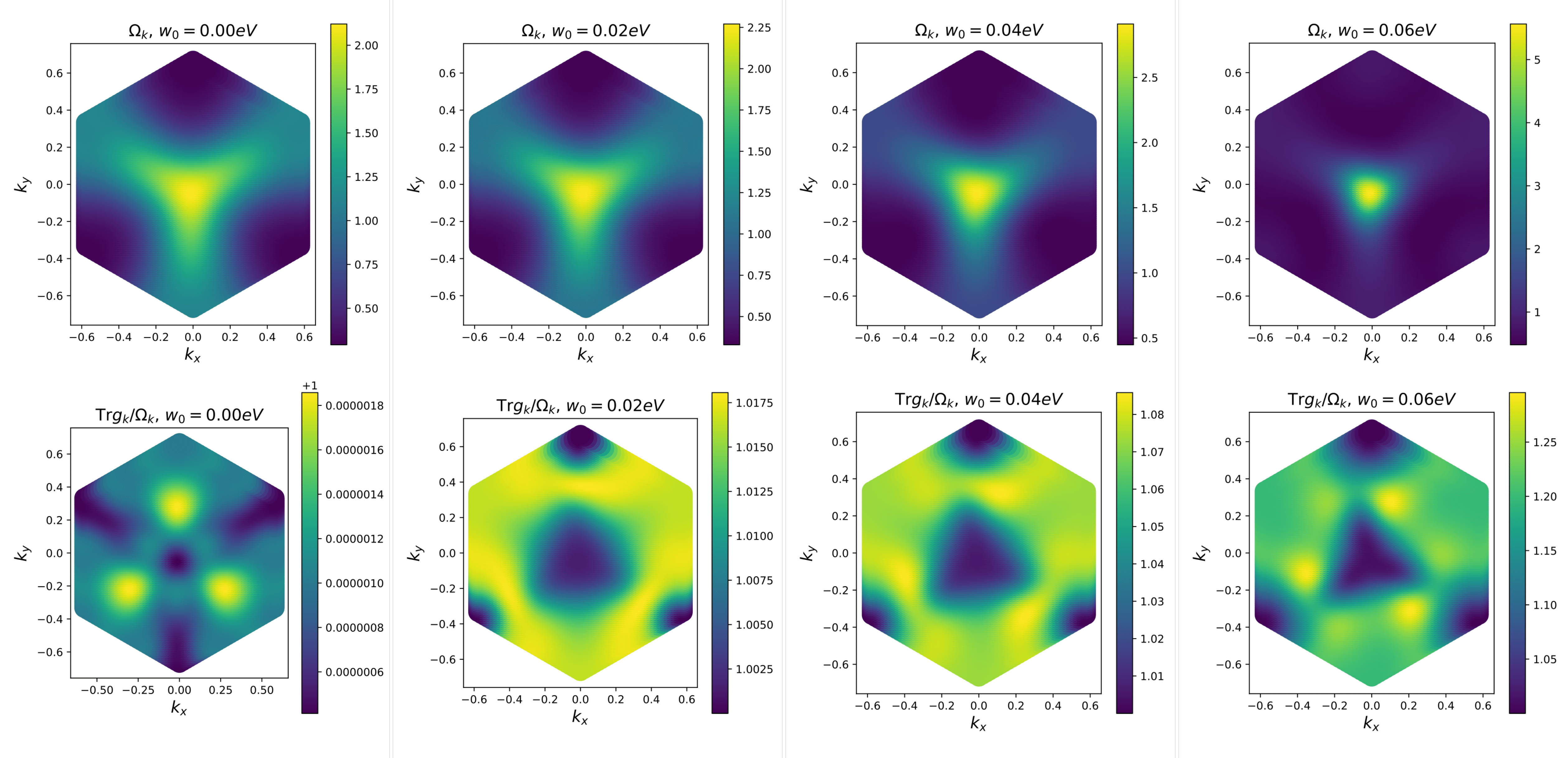}
\caption{Berry curvature and $\text{Tr}g_{\bm k}/\Omega_{\bm k}$ over the Brillouin zone for various $w_0$.}\label{sup:omega_with_w0}
\end{figure}

To gain more insights about the band geometry under realistic parameters, we plot in Fig.~\ref{sup:omega_with_w0} the distribution of Berry curvature and ${\rm Tr}g_{\bm k}/\Omega_{\bm k}$ as a function of nonzero $w_0$ while retaining $t_3=0$ (as we see from Fig.~\ref{sup:var_w0_t3}, $w_0$ is more important in tuning band geometries). The Berry curvature becomes more and more concentrated at the Brillouin zone center, while the trace condition on the other side is less modified near the center. Still, we see over certain regime of $w_0\leq 0.04\rm{eV}$, the band geometry does not deviate too much from the chiral limit.

\begin{figure}
\centering
\includegraphics[width=\linewidth]{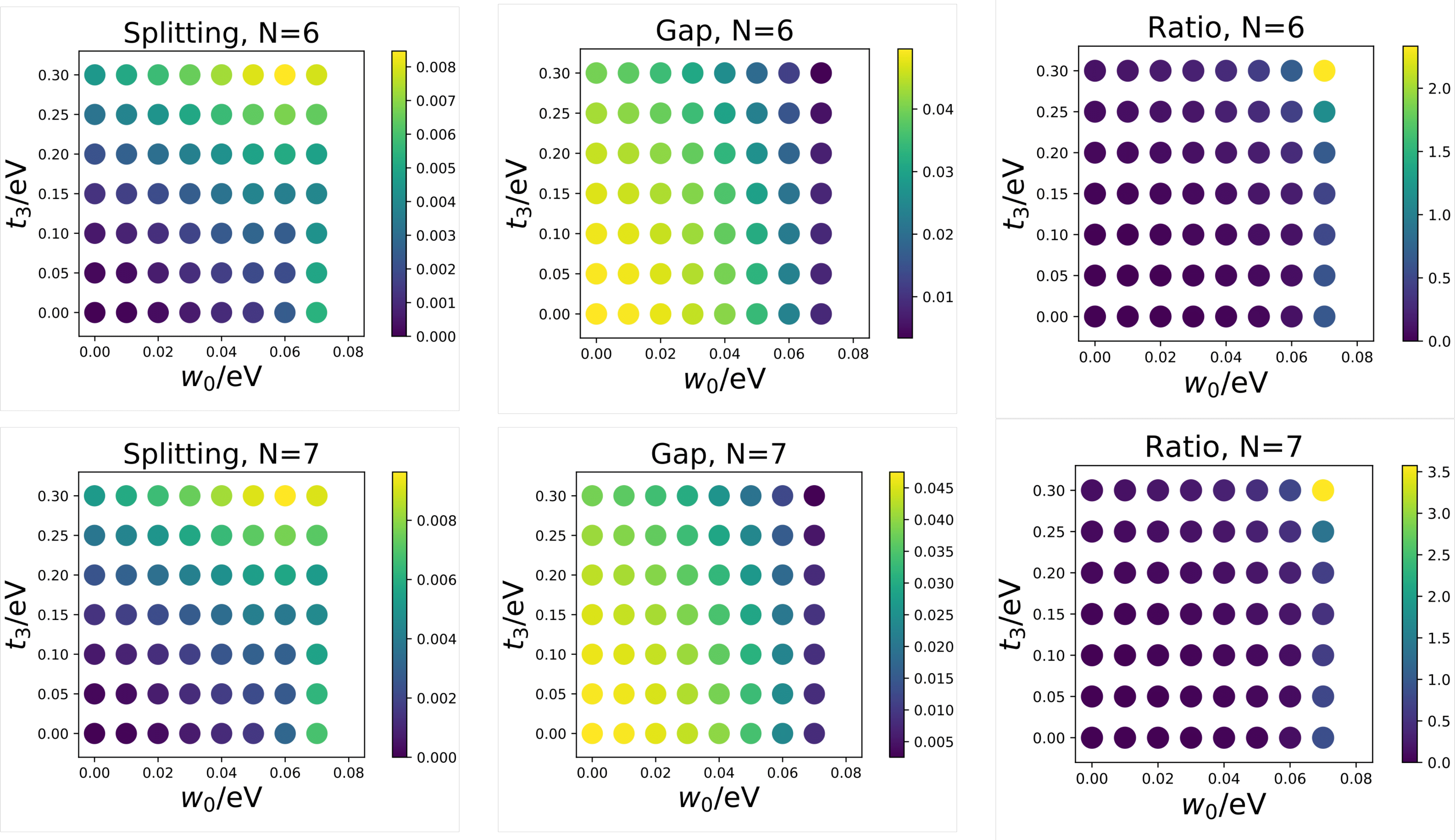}
\caption{Stability of FCIs in the realistic parameter space of TDBG at $\nu=1/5$ filling of the first conduction band. We use the short-ranged $v_1$ interaction. The splitting, gap, and ratio are defined in the text.}\label{sup:manybody}
\end{figure}

In the end, we directly test the stability of FCIs in TDBG beyond the chiral limit. In the main text, we found that in the chiral limit there are exotic model FCIs stabilized by the short-ranged $v_1$ interaction at $\nu=1/5$ filling of one of the two ideal flatbands at charge neutrality. These FCIs are exactly degenerate at zero energy and have an infinite entanglement gap, which is unusual in lattice systems. However, when the system is tuned away from the chiral limit, the ground state could undergo a transition from the FCI phase to a trivial phase. In the following, we study this possibility by considering the $\nu=1/5$ filled first conduction band above the charge neutrality and continuing to use the $v_1$ interaction. We first scan in the $(w_0,t_3)$ parameter space to examine whether the lowest five eigenstates of the interaction Hamiltonian are in the momentum sectors expected for FCIs. If so, we further calculate the ground-state splitting $\Delta_{s}$ (defined as the highest energy minus the smallest energy of the five ground states), the ground-state gap $\Delta_{g}$ (defined as the smallest energy of excited states minus the highest energy of the five ground states), and the ratio $\Delta_{s}/\Delta_{g}$. The results are shown in Fig.~\ref{sup:manybody} for $N=6$ and $N=7$ electrons on the $5\times N$ lattice. We find over a wide range of parameters with $w_0\leq 0.05\rm{eV}$ and $t_3\leq 0.25\rm{eV}$, the ground states are in the FCI momentum sectors, and the ratio $\Delta_{s}/\Delta_{g}$ remains quite small, indicating a well defined separation between the FCI ground-state manifold and higher excited states. Therefore, the FCI phase we find in the chiral limit remains robust in this regime. By comparing Fig.~\ref{sup:var_w0_t3} and Fig.~\ref{sup:manybody}, we find the collapse of the FCI phase coincides with the significant deviation of the band geometry from the ideal case. It would be interesting to study the nature of the ground state after the FCI phase collapses.

The values of $w_0$ and $t_3$ at the border of the FCI phase seem to be close to their values in the realistic TDBG model, where $w_0\approx 0.07-0.08{\rm eV}$ and $t_3\approx 0.28 \rm{eV}$~\cite{PhysRevB.89.035405,Lee:2019aa}. It is very important to take into account more parameters of the realistic TDBG model (such as the particle-hole asymmetry terms and the vertical bias voltage), the dispersion of the active band, and the realistic Coulomb interaction to thoroughly explore the whole phase diagram of the interacting problem. We will leave these studies to future works.

\end{document}